\documentclass[11pt,a4paper]{article}
\pdfoutput=1

\usepackage{jheppub}
\usepackage[utf8]{inputenc}
\usepackage{graphicx}
\usepackage{verbatim}
\usepackage{epsfig}
\usepackage{subfigure}
\usepackage{color}
\usepackage{multirow}
\usepackage{comment}
\usepackage{slashed}
\usepackage{amsmath}
\usepackage{ulem}

\usepackage{amssymb,amsmath,amsfonts,mathrsfs}
\usepackage[dvipsnames]{xcolor}
\usepackage{xcolor}
\usepackage{graphicx}
\usepackage{longtable}
\usepackage{verbatim}
\usepackage{color}
\usepackage[mathscr]{euscript}
\usepackage{framed}
\usepackage{mdframed}  

\usepackage{cancel}
\usepackage{color}
\usepackage{hyperref}
\hypersetup{colorlinks, citecolor=bluscuro, linkcolor=black, urlcolor=bluscuro}
\definecolor{rossos}{cmyk}{0,1,1,0.55}
\definecolor{bluscuro}{rgb}{0.15, 0.2, .85}
\definecolor{bluchiaro}{cmyk}{1,.3,0.,0.1}
\graphicspath{{./Figures/}}

\numberwithin{equation}{section}

\usepackage{tikz}

\usepackage{tcolorbox}
\newcommand{\be}{\begin{equation}\begin{aligned}}
\newcommand{\ee}{\end{aligned}\end{equation}}

\newcommand{\bbe}{\begin{align}}
\newcommand{\eee}{\end{align}}

\newcommand{\bea}{\begin{eqnarray}}
\newcommand{\eea}{\end{eqnarray}}

\def\beq{\begin{equation}}
\def\eeq{\end{equation}}

\def\beqa{\begin{eqnarray}}

	\def\eeqa{\end{eqnarray}}
	
\def\lsim{\mathrel{\rlap{\lower4pt\hbox{\hskip0.5pt$\sim$}}
		\raise1pt\hbox{$<$}}}         %less than or approx. symbol
\def\gsim{\mathrel{\rlap{\lower4pt\hbox{\hskip0.5pt$\sim$}}
		\raise1pt\hbox{$>$}}}         %greater than or approx. symbol

\def\eeqa{\end{eqnarray}}

\def\bq{\begin{quote}}
\def\eq{\end{quote}}

 at 10truept
\newcommand{\arXiv}[2]{\href{http://arxiv.org/pdf/#1}{{\tt [#2/#1]}}}
\newcommand{\arXivold}[1]{\href{http://arxiv.org/pdf/#1}{{\tt [#1]}}}

\def\eq#1{eq.~(\ref{#1})}
\def\fig#1{fig.~\ref{#1}}

\def\M{M_0}
\def\g{y}
\def\V{v}
\def\h{v}

\preprint{CERN-TH-2019-114}
\title{The Selfish Higgs}

\author[a]{G.F.~Giudice,}
\author[b]{A.~Kehagias,}
\author[a,c]{A.~Riotto}

\affiliation[a]{CERN,
Theoretical Physics Department, Geneva, Switzerland.}

\affiliation[b]{Physics Division, National Technical University of Athens\\ 15780 Zografou Campus, Athens, Greece.}

\affiliation[c]{
	Department of Theoretical Physics and Center for Astroparticle Physics (CAP)\\
			24 quai E. Ansermet, CH-1211 Geneva 4, Switzerland.}

\abstract{We propose a mechanism to solve the Higgs naturalness problem through a cosmological selection process. The discharging of excited field configurations through membrane nucleation leads to discrete jumps of the cosmological constant and the Higgs mass, which vary in a correlated way. The resulting multitude of universes are all empty, except for those in which the cosmological constant and the Higgs mass are both nearly vanishing. Only under these critical conditions can inflation be activated and create a non-empty universe.}

\emailAdd{gian.giudice@cern.ch}
\emailAdd{kehagias@central.ntua.gr}
\emailAdd{antonio.riotto@unige.ch}

\begin{document}

\maketitle
\flushbottom

%\newpage
\section{Introduction}

Higgs naturalness remains an open problem in particle physics. The problem has only become sharper after the first runs of the LHC.
On one side, the LHC has identified a scalar particle (which is the origin of the problem) as the agent that completes the electroweak (EW) breaking mechanism at short distances; on the other side, it has widened the gap between the EW scale and the scale of possible new physics. Data from the LHC have amplified the problem, but have not offered a solution yet.

Stimulated by these experimental results, theorists have widened the range of their explorations. One new direction that has been pursued is explaining Higgs naturalness with a selection process during the cosmological evolution. Several interesting examples of this approach have been put forward in the literature~\cite{gia1,gia2,Graham:2015cka,Arkani-Hamed:2016rle,Arvanitaki:2016xds,ib,Geller:2018xvz,Cheung:2018xnu}. Here we will propose a novel mechanism. Our study starts from the following two observations that are generic in the context of cosmological selection.

In traditional solutions of Higgs naturalness based on weak-scale dynamics (technicolor, supersymmetry, composite Higgs, etc.), it is common to disregard the analogous naturalness problem of the cosmological constant. This may be viewed as an acceptable working hypothesis because one can always postulate that the dynamics of the Higgs and the cosmological constant are completely unrelated. However, this hypothesis is hardly defendable in the context of cosmological selection solutions, which generally involve a landscape of values for the Higgs mass. Whatever makes the Higgs mass scan almost necessarily contributes to the energy density of the system and therefore the cosmological constant must scan as well. Thus, our first observation is that {\it any solution to Higgs naturalness based on cosmological selection must simultaneously address the problem of the cosmological constant}. In this paper, we will attempt to construct a mechanism that links the two puzzles.

The second observation, generic of mechanisms based on cosmological selection, is again related to the dynamical variation of theory parameters or, in other words, to the landscape. {\it The presence of a dynamical landscape not only offers a natural setup for anthropic arguments, but makes statistical or environmental considerations almost unavoidable.} For this reason, we will not shy away from relying on anthropic arguments in the construction of our mechanism. When dealing with selection mechanisms based on cosmological evolution, anthropic arguments look as motivated as natural selection in biological evolution. To emphasise the similarity with the role of evolution in biology~\cite{dawkins} we will call our mechanism {\it Selfish Higgs}, since the Higgs acts as an anthropic selector for the emergence of a fairly unique non-empty universe. 

The Selfish Higgs is based on the familiar Standard Model (SM) with the addition of a single non-dynamical field described by a four-form and an inflaton. When coupled to gravity, the four-form contributes to the vacuum energy and it has been proposed in the past as a way of addressing the cosmological constant problem~\cite{ANT,Witten-4,HT,BT1,BT2,duff,BP,saltatory,Gia0,FKRR}, inflation~\cite{Freese:2006fk,K1,K2,K3,Dudas}, and the strong CP problem~\cite{Gia1,Gia2}. In the Selfish Higgs, the four-form is coupled to the SM Higgs and different configurations of the non-dynamical field correspond to different values of the Higgs mass, in a way similar to the original proposal of ref.~\cite{gia1,gia2} and to the models of ref.~\cite{ib}.

We assume that the four-form starts in the early universe in a highly excited state (labelled by a discrete integer $n$), possibly taking different values in different spacetime patches.
The four-form derives from a three-form (the analogous of the potential for an electromagnetic field), which is coupled to (2+1)-dimensional membranes of charge $e$. The charge $e$ has dimension of mass squared and we choose it to define the weak scale. It is technically natural to take it hierarchically smaller than the cutoff mass, which could be in principle as large as the Planck mass. Just like a very intense electric field can be gradually discharged by spontaneous creation of electron-positron Schwinger pairs, so the excited four-form can reduce its energy by creating membranes. In the process, which is governed by quantum tunnelling, the four-form undergoes a configuration transition $n \to n-1$ and a membrane of charge $e$ is nucleated. As a result of this random cascade evolution, in different parts of the universe the Higgs mass squared parameter and the cosmological constant gradually decrease, in a correlated way. The process typically comes to a halt when the cosmological constant has been neutralised. The details of how the process stops are not important for our mechanism. This is because, at the final stages, the tunnelling rate  is so slow that the lifetime of each four-form configuration is much longer than the age of our observable universe, allowing for the possibility that we live in a metastable universe.

The selection criterion that singles out our universe among the multitude of possibilities is purely anthropic, but rather mild: a universe can be `non-empty' only when the cosmological constant and the Higgs mass are close to critical values around zero. The result that the cosmological constant must lie within a small interval around zero follows from Weinberg's well-known considerations~\cite{Weinberg:1987dv}. The novel ingredient of the Selfish Higgs is the feature that only a Higgs near the critical point for EW breaking is capable of igniting the start of inflation by driving the inflaton field away from its true minimum. There are various ways of achieving such a phenomenon and we will present two examples. 

The first mechanism is based on thermal effects due to particle production coming  from the non-adiabatic changes of the Higgs mass squared as the branes sweep the universe. This process is exponentially suppressed whenever the Higgs mass squared (in absolute value) is much larger than the brane charge $e$. The hypothesis that the size of $e$ is in the hundreds-of-GeV range leads to the conclusion that the only non-empty universes must be those in which the Higgs parameters are close to the critical point for EW breaking. The second mechanism is based on a class of inflationary models in which the inflaton is locked at the vacuum in the EW broken phase, while it is free to fluctuate in the EW unbroken phase but without changing the energy density. Inflation is activated only at the critical point between the two phases.

\section{The landscape from a four-form coupled to the Higgs}
\subsection{The action}
The new element that we are adding to the SM is a totally antisymmetric three-form field 
$A_{\mu\nu\rho}$ with a four-form field strength\footnote{Throughout this paper we choose the metric signature $(-,+,+,+)$.} 
\begin{eqnarray}
F_{\mu\nu\rho\sigma}=\partial_\mu A_{\nu\rho\sigma}-\partial_\sigma
A_{\mu\nu\rho}+\partial_\rho A_{\sigma\mu\nu}-\partial_\nu A_{\rho\sigma\mu}\, . \label{FA}
\end{eqnarray}
The dynamics of $A_{\mu\nu\rho}$  is described by the action  \cite{BT1,BT2}
\begin{eqnarray}
S_A=-\frac{1}{48}\int {\rm d}^4 x\sqrt{-g}\,  F_{\mu\nu\rho\sigma}F^{\mu\nu\rho\sigma}
+
S_{\rm mb}\, .  \label{kin4}
\end{eqnarray}
The term $S_{\rm mb}$ describes the coupling of the three-form field to the world-volume of a 3D membrane 
\begin{eqnarray}
S_{\rm mb}=-T \int {\rm d}^3\xi \sqrt{-g^{(3)}}+ \frac{e}{6}\int {\rm d}^3\xi~A_{\mu\nu\rho}
 \frac{\partial x^\mu}{\partial \xi^a}
 \frac{\partial x^\nu}{\partial \xi^b}
 \frac{\partial x^\rho}{\partial \xi^c} \epsilon^{abc}\, . \label{smem}
 \end{eqnarray}
Here $\xi^a$ ($a,b,c=0,1,2)$ are the  membrane coordinates embedded in four-dimensions as $x^\mu(\xi)$; $g^{(3)}_{ab}$ is the induced metric on the membrane; $T$ and  $e$ are the tension and charge of the membrane with mass dimensions three and two, respectively. 
 
The membranes behave as fundamental objects in the low-energy effective theory, valid below $M_{\rm cutoff}$. 
As a result, their structure cannot be resolved and their world-volume is described by a  (2+1)-dimensional field theory with a cut-off of order $M_{\rm cutoff}$. 
The brane world-volume has a vacuum energy that renormalises the brane tension, making it naturally of the size of the cutoff scale,  $T \sim M_{\rm cutoff}^3$.  

On the other hand, an essential point is that the size of the brane charge can be made much smaller than the cutoff  and we take the corresponding mass to define the weak scale ($e  \ll M_{\rm cutoff}^2$). This is a technically natural choice because, in the limit $e\to 0$,  branes are not nucleated, the four-form configuration is globally defined  and  full 4D Poincar\'e invariance is recovered.
Another way of looking at the result is that there are no fields in the world-volume theory that can renormalise the charge $e$. Indeed, the moduli fields parametrising the position of the membrane, which in principle might renormalise the membrane charge, decouple as they are infinitely massive.\footnote{We thank C. Bachas for correspondence about this point.}
The membrane charge can be renormalised if there is vacuum screening because of other charges in the bulk (for instance two-forms), which is however not the case we contemplate. 

The action $S_A$  is invariant under the 
 gauge transformations 
\begin{eqnarray}
A_{\mu\nu\rho}\to A_{\mu\nu\rho}+\partial_\mu B_{\nu\rho}+\partial_\rho
B_{\mu\nu}+\partial_\nu B_{\rho\mu}\, ,
\label{gaugeg}
\end{eqnarray}
where the gauge parameter is a two-form $B_{\mu\nu}=-B_{\nu\mu}$. Because of this gauge redundancy, the three-form $A_{\mu\nu\rho}$ does not contain any propagating degree of freedom, but can have a non-trivial background value such that 
\begin{eqnarray}
F_{\mu\nu\rho\sigma} = f \epsilon_{\mu\nu\rho\sigma} \, ,
\label{deff}
\end{eqnarray}
where $f$ is a constant field in the absence of any source.

The only possible renormalisable interaction between the four-form and the SM fields, invariant under the gauge transformation (\ref{gaugeg}), is through the Higgs portal. The potential of the SM Higgs doublet $H$ then becomes
\begin{eqnarray}
 V_H=-\left(\M^2+\frac{\g}{24} \epsilon^{\mu\nu\rho\sigma}F_{\mu\nu\rho\sigma}\right)|H|^2+\lambda |H|^4 \, ,
 \label{pot}
 \end{eqnarray} 
where $\g$ is a dimensionless coupling and $\M$ is a mass parameter taken to be of the order of the ultraviolet cutoff, in concordance with the naturalness principle ($\M\sim M_{\rm cutoff}$). We have made the discrete choice of a negative bare mass squared for the Higgs. 

The total action of the theory is
\begin{eqnarray}
S= S_A + S_{\rm SM} + S_{\rm b} + S_{\rm grav} \, ,
\label{totalS}
\end{eqnarray}
where $S_A$ is given in \eq{kin4} and
$S_{\rm SM}$ is the usual SM action augmented with the four-form coupling to the Higgs as in \eq{pot}
\begin{eqnarray}
S_{\rm SM}=\int {\rm d}^4 x\sqrt{-g}\left(
-D_\mu H^\dagger D^\mu H -V_H+\cdots\right) \, .
\end{eqnarray}
$S_{\rm b}$ is a pure boundary term 
 \begin{eqnarray}
S_{\rm b}=  \frac16 \int {\rm d}^4 x\, \partial_\mu \left[ \sqrt{-g} \left( F^{\mu\nu\rho\sigma}A_{\nu\rho\sigma}-{\g}\,\epsilon^{\mu\nu\rho\sigma}A_{\nu\rho\sigma}|H|^2\right) \right] \, , 
\label{Sb}
 \end{eqnarray}
  which, although it does not contribute to the classical equations, removes an inconsistency between field equations and the on-shell action \cite{duff,Duncan:1989ug} and leads to stationary action under variations that leave the four-form fixed at the boundary. Note that, after integrating out the Higgs at one loop, the two terms proportional to $y$ in eqs.~(\ref{pot}) and (\ref{Sb}) exactly cancel, showing explicitly that the coupling $y$ cannot renormalise the brane charge $e$.

Finally, $S_{\rm grav}$ is the gravitational action, which is relevant for the tunnelling dynamics across membranes, 
\begin{eqnarray}
S_{\rm grav} =  
\int {\rm d}^4 x\sqrt{-g} \, \left(\frac{M_P^2}{2}\, R+\Lambda_{0}\right)
+
 \oint {\rm d}^3x \sqrt{-h}\, M_P^2\, K \, .
 \label{grav}
 \end{eqnarray}
 Here $M_P$ is the reduced Planck mass and $-\Lambda_0$ is the bare cosmological constant. We choose $\Lambda_0$ to be positive (corresponding to a negative cosmological constant) with size of the order of the cutoff scale size, as required by naturalness ($\Lambda_0\sim M_{\rm cutoff}^4$). The last term of \eq{grav} is the Gibbons-Hawking term~\cite{Gibbons:1976ue} where the surface integral is over spacetime boundaries, with
 $h_{ab}$ being the induced metric and $K=K^a_a$ the trace of the extrinsic curvature $K_{ab}$. The theory is manifestly CP-conserving, up to the usual SM sources of CP violation.
 
 \subsection{The field equations}
 
In order to study the dynamics of the system we consider the field equations expressed in terms of the real scalar neutral Higgs component $h$, defined such as $H=(0, h/\sqrt{2})$, and the field $f$ defined in \eq{deff}. For simplicity, we turn off gravity (which plays no role in this context) and varying the action in \eq{totalS} we obtain 
\begin{eqnarray}
&&\epsilon^{\mu\nu\rho\sigma}\partial_\mu\Big{(}f-\frac{\g}{2} h^2\Big{)}=-e \int {\rm d}^3\xi~\delta^4\left( x-x(\xi)\right) \,
 \frac{\partial x^\nu}{\partial \xi^a}
 \frac{\partial x^\rho}{\partial \xi^b}
 \frac{\partial x^\sigma}{\partial \xi^c} \epsilon^{abc} \, , \label{fe11}\\
 &&
\Box h =\left(-\M^2+\g f\right) h+\lambda h^3 \, . \label{fe12}
\end{eqnarray}
Let us focus on the constant-field vacuum configuration
\be
\langle h \rangle = v \, .
\ee
From \eq{fe11} we learn that, when a membrane is nucleated, 
$f$ and $\h$ are constant on both sides of the membrane wall, but there  is a jump in their values across the wall
\begin{eqnarray}
\Delta f-\frac{\g}{2} \Delta \h^2=e\, . \label{Df}
\end{eqnarray}
As argued in ref.~\cite{BP}, not only the background field makes discrete jumps across membranes, but its value is actually quantised in units of $e$ 
\begin{eqnarray}
f-\frac{\g}{2} \h^2=e\, n\, , ~~~n\in \mathbb{Z} \, .
\label{quants}
\end{eqnarray}
This quantisation is expected whenever the theory has a UV completion such as string theory, in which gauge fields and their duals are related. The Dirac quantisation condition of magnetic sources then implies quantisation of electric sources, whose role in our case is played by the constant value of the four-form.

Once we accept the quantisation condition, we find a discrete (and infinite) set of possible background solutions labelled by the positive integer $n$ 
\begin{eqnarray}
\begin{cases} f= e\, n \\ \h^2 = 0
\end{cases}
~~~{\rm for}~n> n_c ~~~\mbox{(unbroken phase)}\, ,
\label{bck1}
\end{eqnarray}
and
\begin{eqnarray}
\begin{cases} f= \frac{\g \M^2+2\lambda \, e\, n}{\g^2+2\lambda} \\ \h^2 = \frac{2(\M^2-\g \, e\, n)}{\g^2+2\lambda}
\end{cases}
~~{\rm for}~n\le n_c ~~\mbox{(broken phase)}\, ,
\label{bck2}
\end{eqnarray}
where $n_c$ is the largest integer smaller than $\M^2/\g e$.

\subsection{The scanning parameters}

We want to evaluate the effective cosmological constant including the contribution to the vacuum energy from the background fields $f$ and $h$. The energy-momentum tensor for the four-form and the Higgs is
\begin{eqnarray}
T^{\mu\nu}=\frac{1}{6}F^{\mu\kappa\rho\sigma}{F^\nu}_{\kappa\rho\sigma}-\frac{1}{48}g^{\mu\nu}F^{\kappa\rho\sigma\lambda}F_{\kappa\rho\sigma\lambda}+\partial^\mu h\partial^\nu h-g^{\mu\nu}
\left(\frac{1}{2}\partial^\rho h\partial_\rho h+U\right)\, , \label{e-m}
\end{eqnarray}
\begin{eqnarray}
U=-\frac{\M^2}{2} h^2+\frac{\lambda}{4}h^4 \, . \label{U}
\end{eqnarray}
The reason why only the part of the potential that is independent 
of the four-form contributes to the energy-momentum tensor in \eq{e-m} is that the term
\begin{eqnarray}
\sqrt{-g}\, \epsilon^{\mu\nu\rho\sigma} F_{\mu\nu\rho\sigma}\, h^2
\end{eqnarray}
does not depend on the metric. Thus, the energy density on the field
 background is
\begin{eqnarray}
\rho \equiv T^{00} = \frac{f^2}{2} + U \, .\label{rho1}
\end{eqnarray}
Superficially it may seem that the energy-density in \eq{rho1} is independent of the coupling $\g$.  However, the dependence is hidden inside the Higgs contribution to $f$. By writing the quantisation condition as 
$f=f_n+\g h^2/2$ with $f_n =en$, we find  
\begin{eqnarray}
\rho=\frac{f_n^2}{2}-\left(\M^2-\g f_n\right)\frac{h^2}{2}+\left(2\lambda+\g^2\right) \frac{h^4}{8}\, ,
\label{rho2}
\end{eqnarray}
which shows explicitly the dependence on  $\g$.

The cosmological constant $\Lambda$ and the Higgs mass parameter $\mu_H^2$ (defined such that the Higgs potential is $V_H=\mu_H^2 |H|^2 +\lambda |H|^4$) corresponding to
the solutions labelled by the integer $n$ in eqs.~(\ref{bck1})--(\ref{bck2}) are given by
\begin{eqnarray}
\begin{cases} 
\Lambda= -\Lambda_0 +\frac{e^2 n^2}{2} \\
\mu_H^2 =-\M^2 +\g \, e\, n
\end{cases}
~~~{\rm for}~n> n_c ~~\mbox{(unbroken phase)}
\label{bbck1}
\end{eqnarray}
and
\begin{eqnarray}
\begin{cases} 
\Lambda= -\Lambda_0 +\frac{\lambda e^2n^2 +\g \M^2e\, n-\M^4/2}{\g^2+2\lambda} \\
\mu_H^2 =\frac{2\lambda (\g \, e\, n -\M^2)}{\g^2+2\lambda}
\end{cases}
~~~{\rm for}~n\le n_c ~~\mbox{(broken phase)}\, .
\label{bbck2}
\end{eqnarray}
The important observation for the Selfish Higgs is that the coupled system four-form/Higgs leads to a landscape of possible background solutions, on which both the cosmological constant and the Higgs mass parameter vary. However, $\Lambda$ and $\mu_H^2$ do not scan independently, but remain correlated as $n$ changes, as shown in eqs.~(\ref{bbck1})--(\ref{bbck2}). The next step is to study how this landscape can be dynamically populated and this is the subject of the following section.

Before concluding this section, we remark that $\mu_H^2$ scans linearly with $n$ and $\Lambda$ scans quadratically, as shown in eqs.~(\ref{bbck1})--(\ref{bbck2}). This behaviour of $\Lambda$  is the cause of the `gap problem' in the Brown-Teitelboim mechanism for the cosmological constant relaxation. Indeed, the splitting in $\Lambda$ between two contiguous states around the value of $n$ which neutralises the cosmological constant is
\begin{eqnarray}
\Delta \Lambda \equiv \left[ \Lambda (n+1)- \Lambda (n) \right]_{\Lambda =0} \approx \left[ e^2 n \right]_{\Lambda =0} \approx e \sqrt{2\Lambda_0} \, .
\label{ccscan}
\end{eqnarray}
Since $\Lambda_0$ is naturally at the cutoff, getting a sufficiently fine scanning of the cosmological constant requires a phenomenally small value of the brane charge. One must choose $e\sim (10^{-42}~{\rm GeV})^2$, which defines the `gap problem'. 

On the other hand, due to the linear dependence on $n$, the scanning of $\mu_H^2$ occurs through uniform steps
\begin{eqnarray}
\Delta \mu_H^2 \equiv \mu_H^2 (n+1)- \mu_H^2 (n) = \g \, e \, ,
\end{eqnarray}
which do not involve quantities parametrically equal to the cutoff scale. A crucial assumption of the Selfish Higgs is that these steps are of the order of the weak scale
\be
\g \, e = {\mathcal O} (m_h^2) \, ,
\ee
where $m_h=125$~GeV is the physical Higgs boson mass.

\section{The Selfish Higgs as an anthropic selector}

\begin{figure}[t]
    \begin{center}
      \includegraphics[scale=0.3,angle=0]{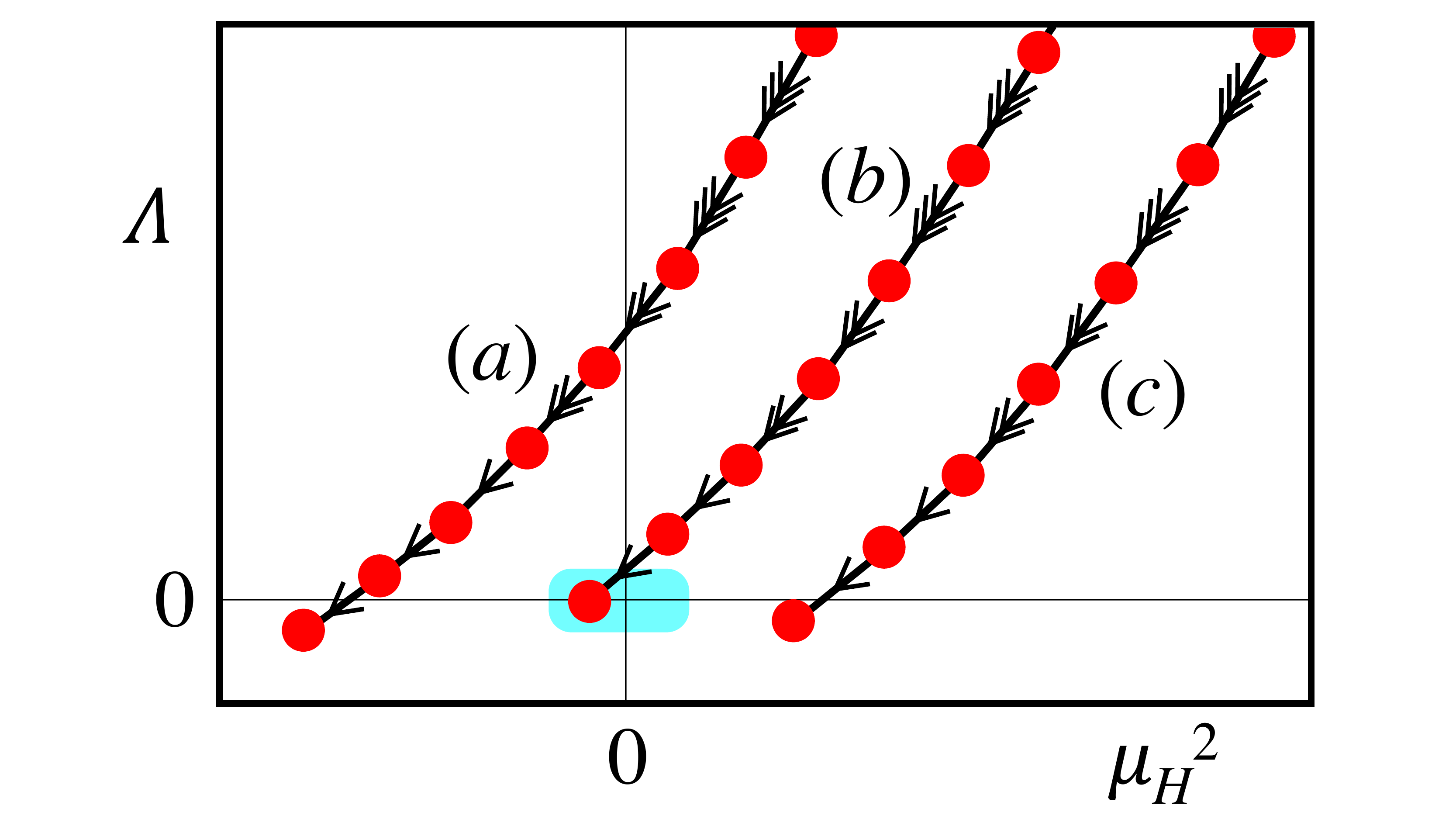}
    \end{center}
     \caption{\small Sketch of possible trajectories in the plane of cosmological constant ($\Lambda$) and Higgs mass squared parameter ($\mu_H^2$) coming from the evolution of four-form configurations. Transitions become exponentially slow as $\Lambda$ is reduced and typically come to a halt after the first jump into AdS or Minkowski. The condition $\Lambda \approx 0$ can be reached with EW symmetry in the broken phase ($a$), unbroken phase ($c$), or near-critical ($b$). The coloured region shows the area selected by the Selfish Higgs.}  
     \label{fig:traj}
\end{figure}

\subsection{Scanning the Higgs mass with spontaneous membrane nucleation}

As previously shown, the four-form has no dynamics at the classical level, although it can attain different field configurations labelled by the integer $n$. Quantum mechanically, these configurations are unstable and can tunnel into each other through non-perturbative effects. Starting from a spacetime region with charge $n$, a membrane can be spontaneously created encompassing a subregion in which the field is in the $(n-1)$ configuration. The energy stored in the membrane tension is taken from the energy gain of lowering the field by one charge unit. The membrane will then expand at the speed of light, while new membranes can nucleate in its interior. The tunnelling rates between different configurations have been studied in ref.~\cite{BT1,BT2} (see also ref.~\cite{quevedo}). The analysis shows that the rate for lowering $n$ is overwhelmingly larger than for transitions with increasing $n$, which can happen through gravitational instantons. 

We imagine an initial condition in which the four-form is in highly excited states throughout the universe with
 large values of $n$ in various spacetime patches. This corresponds to an initial condition with large and positive $\Lambda$ and $\mu_H^2$. 
 As a result of the tunnelling dynamics, the four-form will gradually discharge randomly in different parts of the universe, 
 effectively creating a multiverse in which $\Lambda$ and $\mu_H^2$ scan in a correlated way, according to eqs.~(\ref{bbck1})--(\ref{bbck2}). Through discrete jumps, the values of $\Lambda$ and $\mu_H^2$ will progressively decrease along one of the trajectories shown in \fig{fig:traj}. The values of $\M$ and $\Lambda_0$ select the trajectory. Along trajectories of type $(a)$ EW breaking occurs while the system is still in dS space. For type $(c)$ the system always remains in the unbroken phase, even when the cosmological constant crosses zero. Type $(b)$ corresponds to the special coincidence in which $\Lambda$ and $\mu_H^2$ nearly vanish simultaneously.
 
 As argued in ref.~\cite{BT1,BT2}, the tumbling process terminates when the cosmological constant has been neutralised and the field configuration has quantum tunnelled into Minkowski or has made its first jump into AdS space. This happens as long as the brane tension $T$ is sufficiently large, corresponding to a condition that can be easily satisfied in our case. However, this condition is not strictly necessary for our mechanism. Indeed, for us it is sufficient that a field configuration which supports a non-empty universe has a lifetime longer than the present age of our universe. This is certainly the case because the tunnelling probability between two consecutive configurations in dS space is given by
 \be
 {\mathcal P}(n+1 \to n)\approx \exp \left( - \frac{24\pi^2M_P^4}{\Lambda_{n+1}} \right) \, ,
 \ee
for $\Lambda_{n+1} \ll T^2/M_P^2$. At the last stages of evolution, when the effective cosmological constant is very small,
 the tunnelling transitions are extremely slow and all corresponding universes have a viable metastable nature. So further transitions into AdS (even if they occur) do not invalidate our mechanism.
 
Equation~(\ref{ccscan}) shows that, for values of the unit charge relevant to our study ($\g e \sim m_h^2$) the cosmological constant is not scanned finely enough to have a reasonable probability of landing within the observed range. Therefore, we will assume that the two dimensionful parameters $\M$ and $\Lambda_0$ scan continuously in the UV theory within a range
around $M_{\rm cutoff}$, as a consequence of being functions of some underlying dynamical variables. The parameter $e$, which is likely to have a different origin since its mass scale is much smaller than $M_{\rm cutoff}$, is held fixed. Spacetime patches that start with different values of the bare parameters $\M$ and $\Lambda_0$ correspond to different trajectories in \fig{fig:traj} and to different offsets of their discrete points. There will be cases in which one particular configuration lands arbitrarily close to $\Lambda=0$. 

The scanning of $\M$ and $\Lambda_0$ opens up the possibility of tunnelling between different trajectories of \fig{fig:traj}. The corresponding transition rates depend on unknown physics at the cutoff, but are likely to be suppressed with respect to transitions along a single trajectory. At any rate, these transitions would not modify the general features of the Selfish Higgs mechanism. Another possible concern could be the Higgs tunnelling into the large-field configurations that minimise the energy in the SM when the Higgs quartic coupling becomes negative~\cite{instab1,instab2}. However, the EW vacuum is generally not destabilised during the brane evolution because, as we will show in the following, the size of $\mu_H^2$ is typically larger than the Hubble parameter $H^2$ and Higgs fluctuations in dS are damped.

\subsection{Selecting a non-empty universe with thermal effects}

The selection condition that we employ for the Selfish Higgs is based on anthropic considerations, but is very minimalistic: the existence of a non-empty universe containing sufficient entropy density for a sufficiently long time. 

The vast majority of the configurations in our dynamical landscape corresponds to empty (or nearly empty) dS universes, in which the exponential expansion dilutes any possible initial particle content. Our selection condition then restricts the possible universes to a narrow band around $\Lambda =0$. A quantitative way of assessing the anthropic range of the cosmological constant, based on structure formation, has been famously proposed by Weinberg~\cite{Weinberg:1987dv} and we will adopt it here. Having narrowed down the values of $\Lambda$ is not sufficient to satisfy the selection criterion because the theory as such still seems to predict empty universes. We need to identify the source that ignites a period of inflation followed by a radiation epoch, giving rise to the universe as we observe it. We also want to find the necessary ingredients within the theory of the four-form coupled to the SM. 

As a possible tool provided by the theory to reach our goal, we consider the non-thermal particle production generated by the non-adiabatic dynamics during membrane nucleation. There are two possible sources for these effects.

The first is coming from the sudden jumps of the Hubble rate $H$, as the membrane is nucleated. The discontinuity of $H$ across the membrane wall is 
\begin{eqnarray}
\Delta H^2 = \frac{e^2 n}{3M_P^2} \, ,
\label{cip}
\end{eqnarray}
which, around the critical point for EW breaking $n=n_c$ becomes
\begin{eqnarray}
\Delta H^2 \approx \frac{\M^2}{3yM_P^2} \, e \, .
\label{ciop}
\end{eqnarray}
This is of the order of the weak scale or smaller, if the cutoff scale is below the Planck scale.

Whenever the Higgs is non-minimally coupled to gravity, it acquires a mass in dS space given by
\be
m_h^2=12 \xi \, H^2 \, ,
\label{csi}
\ee
where $\xi$ is the Higgs coupling to the Ricci scalar. Even in the absence of this coupling, if the Higgs is quantum-mechanically excited,  it acquires a one-loop mass of order $H^2$~\cite{chen2}
\be
m_h^2=\sqrt {\frac{6\lambda}{\pi^3}}H^2 \, .
\ee
In the presence of Higgs fluctuations, the coupling to the Higgs feeds one-loop masses for $W$ and $Z$ bosons~\cite{chen2}
\be
m_W^2=\frac{3 g^2 H^4}{8\pi^2 m_h^2} \, , \quad m_Z^2=\frac{3 g^2 H^4}{8\pi^2 \cos^2 \theta_W m_h^2} \, ,
\ee
while the photon, gluon and fermions remain massless. The mass of the gauge boson is induced by the variance of the Higgs, with its VEV being still vanishing in each Hubble volume. This is the reason why fermions do not acquire a mass. 
 
 The presence of $H$-dependent masses may cause some of the particles of the SM to be created at each membrane nucleation as the Hubble rate $H^2$ jumps by units of $e$. In particular, the last jump into flat spacetime would induce gravitational particle production giving rise to significant excitations of such particles  because of the  non-adiabatic evolution of modes. However, this source of particle production does not select small Higgs VEV $\V$ as a special point and thus it is not appropriate for our purposes.
   
 Therefore we work under the assumption that the Higgs fluctuations are exponentially damped, taking $\mu_H^2\gg H^2$. The hypothesis $\mu_H^2\gg H^2$ is actually easily satisfied since we can write (in the unbroken phase)
 \be
 \frac{\mu_H^2}{H^2} = \frac{6\g \, M_P^2}{\M^2} \left( \frac{N-1/\g}{N^2-2\Lambda_0/\M^4}\right)\, ,  \quad N=\frac{e\, n}{\M^2} \, .
 \label{muH}
 \ee
 The dimensionless quantity $N$ has been defined such that it varies in a range of order unity. Thus $\mu_H^2/ H^2$ can be naturally made large (anywhere away from the critical point for EW breaking) by taking the cutoff somewhat lower than the Planck scale. An alternative (and even more efficient) way of eliminating particle production from jumps in $H$ is to assume a conformally invariant coupling of the Higgs to curvature ($\xi=1/6$). We recall that $\xi = 1/6$ is stable under quantum corrections. 
  
We turn to the second source of particle production, which can take place through rapid non-adiabatic changes of the particle mass induced by changes of the Higgs VEV. We illustrate the effect for a generic scalar particle $P$ whose mass $m_P$ is $v$-dependent, and we explain how our considerations can be extended to vector or fermion particles.
The starting point is the mode decomposition of the real scalar field $P(x)$ in a curved space with Friedmann-Robertson-Walker metric~\cite{pt}
\be
P(x) = \int \frac{{\rm d}^3k}{(2\pi)^{3/2}}\, a(\eta)\left[ a_k f_k(\eta) e^{i\vec{k} \cdot \vec{x}}+a_k^\dagger f^*_k(\eta) e^{-i\vec{k} \cdot \vec{x}}\right] \, ,
\ee
where $a$ is the scale factor and $\eta$ is conformal time. The equation of motion of the momentum modes $f_k$ is
 \be
 f_k'' + \omega^2 f_k =0 \, \quad \omega^2 = {\vec k}^2 +m_P^2\, a^2 + \left( \xi -\frac 16 \right) a^2 R \, ,
 \label{mode}
 \ee
where primes denote derivatives with respect to conformal time and $R=6a''/a^3$. As expected, for $\xi =1/6$ the scalar particle is insensitive to fluctuations of the curvature due to changes in the Hubble rate. Here we are focusing on changes in the mass $m_P$ from jumps in $v$. We remark that, by dropping the last term in \eq{mode} ({\it i.e.} by setting $\xi =1/6$), our analysis is exactly valid also for the transverse modes of vector bosons. The treatment of the longitudinal modes of vector bosons and fermions requires extra terms in \eq{mode}~\cite{pt}, but the extension is straightforward and leads to results similar to those presented here for the scalar case.

Consider the nucleation of a membrane separating a spacetime region in which the four-form is in the configuration $n$, created out of a region with $n+1$. The momentum modes associated with frequencies before ($\omega_{n+1}$) and after ($\omega_{n}$) the transition are related by a Bogoliubov transformation $f_{k}^{(n+1)}=\alpha_k f_{k}^{(n)}+\beta_k f_{k}^{(n)*}$. 
 The occupation number of a mode ($n_k= |\beta_k|^2$) is an adiabatic invariant, and particle production happens only if a mode  evolves non-adiabatically. The total particle number density is
 \be
 n_P = \frac{N_P}{2\pi^2 a^3} \int {\rm d}k \, k^2 \, n_k \, ,
 \label{nppp}
 \ee
 where $N_P$ is the number of degrees of freedom of the particle $P$.

An analytic computation of the Bogoliubov coefficients can be done assuming the following simple ansatz for the VEV profile in conformal time~\cite{pt} (see also ref. \cite{jump})
\be
 \V^2(\eta)=\frac{\V_{n+1}^2+\V^2_{n}}{2}-\frac{\V^2_{n+1}-\V^2_{n}}{2}\, \tanh\, (\kappa  \eta H_{n+1}) \, ,
 \label{prof}
 \ee
 where $\kappa$ is a parameter controlling the speed of the transition ($\kappa =0$ corresponds to infinitely slow transition and $\kappa \to \infty$ to infinitely fast). In the case of \eq{prof}, one finds that the occupation number of particles $P$ produced during the non-adiabatic change of the VEV is 
 \be
n_k=\frac{\sinh^2\left[\frac{\pi(\omega_{n+1}-\omega_{n})}{2\kappa H_{n+1}}\right]}{\sinh\left(\frac{\pi\omega_{n+1}}{\kappa H_{n+1}}\right)\sinh \left(\frac{\pi\omega_{n}}{\kappa H_{n+1}}\right)}  \, .
\label{sinh}
 \ee

Particles produced by membrane nucleation occurring at large $H$ are irrelevant, because the fast expansion quickly dilutes their density. The only relevant reheating can occur at the last transition, when the universe settles into a configuration with nearly vanishing cosmological constant. For this transition, the corresponding Hubble is $H_{\rm last}^2 \approx e \, \sqrt{2\Lambda_0}/(3M_P^2)$, see \eq{cip}. 

Let us first consider the case $\mu_H^2<0$ (broken phase) in which all SM particles acquire masses proportional to $v$ and are produced non-thermally according to the mechanism described above. We define $m_P =g_P\, v/2$, where the normalisation is chosen such that $g_P$ is equal to the weak coupling $g$ when $P$ is the transverse $W$ boson, and $g_P=(8\lambda )^{1/2}$ when $P$ is the Higgs. In the limit $\V > H_{\rm last}$, the number density of $P$ particles is\footnote{In the limit of large $b\equiv 2\pi m_P / (\kappa H_{n+1})$,  eqs.~(\ref{sinh}) and (\ref{nppp}) become
$$
n_k \!= \!e^{-b\sqrt{1+k^2/m_P^2}}  , ~ n_P\! =\! \frac{N_P m_P^3 e^{-b}}{(2\pi b)^{3/2}} \, I(b)  , ~
I(b) \! \equiv \! \sqrt{\frac{2}{\pi}} \int_0^\infty {\rm  d} x\, x^2  e^{b\left( 1- \sqrt{1+x^2/b}\right)} \! = \! \sqrt{\frac{2b}{\pi}}e^b K_2(b)\!=\! 1 + \mathcal {O} (b^{-1}) . 
$$
Hence, \eq{npar} follows.}
\be
n_P =N_P  \left( \frac{g_P^2 \, v \sqrt{e}}{8\pi c}\right)^{3/2}  \exp \left( - c\, \frac{v}{\sqrt{e}}\right) \, , \quad \quad c\equiv \frac{\sqrt{3}\, \pi  \, g_P\, M_P}{\kappa \, (2\Lambda_0)^{1/4}} \, .
\label{npar}
\ee
Since we expect $c$ to be a number of order unity, \eq{npar} gives a strong exponential suppression  unless $\V$ is of the order of $H_{\rm last} \sim \sqrt{e}$. This suppression is due to the inefficiency of particle production whenever the relative change in $v$ is small. Only for $v^2 ={\mathcal O} (e)$, the change in the Higgs VEV is of the order of the VEV itself, and particle production is efficient. The result can be also understood as a Boltzmann suppression for producing particles with masses of order $v$ from dynamics with Hubble rate $H \sim \sqrt{e}$. 

When $\mu_H^2>0$ (unbroken phase), the Higgs is the only massive particle in the SM, with mass equal to $\mu_H$. Higgs production from non-adiabatic changes of its mass can take place but, analogously to the previous case, it is exponentially suppressed when $\mu_H$ is larger than $\sqrt{e}$. Note that particle production is less efficient in this case, because only the Higgs degrees of freedom are involved, leading to a preference for the broken phase.

In summary, we have found that, out of the multitude of universes generated by the random process of brane nucleation, only those with small $|\Lambda |$ and $|\mu_H^2|$ are `non-empty'. In these special universes, brane nucleation leads to the production of SM particles that will rapidly thermalise creating a bath with weak-scale temperature. However, these universes do not have the right properties to resemble our own for at least two reasons. First, the nucleated bubbles with $\Lambda \approx 0$ will expand and asymptotically fill up a very large fraction of space, but cannot percolate in the expanding dS environment in which they are immersed~\cite{gw}. This is problematic\footnote{We thank M.~Geller and A.~Hook for bringing up this point.} because a single bubble has size $H^{-1}\sim M_P/T^2$ and entropy density $s \sim T^3$, so that its total entropy $S\sim M_P^3/v^3 \sim 10^{48}$ is insufficient to contain our universe, whose present entropy inside the horizon is about $10^{88}$. The second problem is that  the nucleated bubbles do not have the density perturbations needed to seed structure formation. As usual, these problems can be solved with a stage of inflation.

The critical element of the Selfish Higgs is to provide the spark that ignites inflation by displacing the inflaton field. We assume that, throughout the brane evolution, the inflaton field remains anchored at its minimum. This can happen, for instance, because of the mass acquired by the inflaton in dS space from its coupling to gravity, given by $m^2 = CH^2$ where $C$ is a model-dependent constant. As shown in \eq{csi}, for the coupling of the inflaton to curvature one finds $C=12 \xi$. The $H$-dependent mass disappears inside the last bubble where Minkowski space is nucleated. In general, the inflaton vacuum in dS and Minkowski space are different and one may worry about the inflaton first getting displaced and then oscillating around its new minimum, eventually dominating the energy density of the universe. However, as shown in ref.~\cite{Linde:1996cx}, for sufficiently large $C$, the inflaton shifts into its new vacuum by tracking adiabatically the change in the potential and oscillations are exponentially damped. This happens whenever $\xi$ is even only moderately larger than 1. An alternative possibility is that the minimum in dS and Minkowski space is the same because it corresponds to a point of enhanced symmetry~\cite{Dine:1995uk} and the inflaton does not move during the evolution of the brane configurations. 
Under any of these assumptions, thermal effects are the only source able to initiate the inflationary process.

Here we will not try to build specific models, but only describe how a thermal origin for inflation can be fairly generic. To react to temperature effects, the inflaton must have a mass less than the weak scale ($m_I \lsim 10^2$~GeV) and be sufficiently coupled to SM particles: for renormalisable interactions, the coupling must satisfy $g_I \gsim (v/M_P)^{1/2}\sim 10^{-8}$ and, for dimension-5 interactions, the corresponding scale must satisfy $f_I \lsim (vM_P)^{1/2} \sim 10^{10}$~GeV.  There are several ways in which thermal effects can start inflation. The inflaton can be displaced either by thermal fluctuations or because of modifications of the effective potential. As a result, the inflaton can find itself away from the zero-temperature minimum and start a slow-roll evolution. Regions where this happens will quickly dominate due to the exponential expansion. The dynamics and the corresponding spectrum of perturbations from such models of EW-scale inflation have been worked out, for instance, in ref.~\cite{kt}.  

Another possibility is that the inflaton gets trapped in a false vacuum after being pushed by thermal effects~\cite {BP}. Eventually, the inflaton will  tunnel through the barrier, slowly rolling down towards the true minimum of the potential.  The barrier may be created by a renormalisable term of the form $h^2\phi^2$, where $\phi$ is the inflaton field. If the last membrane nucleation creates an unbroken phase, such a barrier term is absent. On the other hand, if the Higgs VEV is too large, thermal effects are not able to push the inflaton field away from the false vacuum. Only if the VEV is of the order of the EW scale, the inflaton may be efficiently pushed towards the true vacuum at the origin. This will lead to the usual process of inflation, with at least 60 e-foldings to attain the required homogeneity, with the creation of density perturbation, and with the final reheating of the universe.

\subsection{Selecting a non-empty universe with inflationary models}

The basic idea of the Selfish Higgs is that, out of the many universes obtained by the simultaneous scanning of the cosmological constant and the Higgs mass, only those with small $\Lambda$ and $\mu_H^2$ can sustain entropy density for a sufficient long time, while all others remain essentially empty. In the previous section we have used particle production during brane nucleation as the agent that activates inflation. However, one can imagine many variations of the selection mechanism. One possibility is to use models with appropriate Higgs-inflaton interactions such that  the inflationary phase is triggered only by a near-critical Higgs mass. 

As an illustration of the idea, let us consider a field $\phi$ which, as long as EW symmetry remains unbroken, behaves as an exact Goldstone boson and thus $V(\phi) =0$. However, the Higgs VEV triggers an explicit breaking of the global symmetry, generating a potential for $\phi$ of the form 
\be
V(\phi) = v^\alpha F(\phi) \, ,
\label{vpot}
\ee
 where the exponent $\alpha$ and the function $F$ are model-dependent quantities. This structure is obtained, for instance, in the case of an axion-like particle whose continuous global symmetry is broken into a discrete shift symmetry by non-perturbative QCD effects that generate a periodic potential with $F(\phi) \sim \cos(\phi/f_\phi)$ proportional to a field condensate. Through the quark-mass dependence, the condensate is proportional to the Higgs VEV $v$, with $\alpha \approx 1$. This setup is at the basis of the relaxion mechanism~\cite{Graham:2015cka} and can be generalised to new high-scale strong interactions, as done in ref.~\cite{Espinosa:2015eda}. While the case of an axion-like particle is an interesting realisation that fulfils our requirements, our considerations apply more generally to any inflationary model with a potential of the form (\ref{vpot}).
 
Let us trace the evolution of the field $\phi$ during the process of brane nucleation, in the class of models defined by \eq{vpot}. As long as the system remains in the EW unbroken phase (along trajectories of type {\it (c)} in \fig{fig:traj}), the field $\phi$ moves randomly but does not activate inflation because $V(\phi)$ exactly vanishes. In this situation, the universe remain empty also when the condition $\Lambda \approx 0$ is reached. 
 
 If EW symmetry is broken before the cosmological constant is neutralised (trajectories of type {\it (a)} in \fig{fig:traj}), then the field $\phi$ rolls into the minimum of its potential, but any entropy production generated in the process is rapidly diluted by the dS expansion. Further brane nucleation does not change the location of the minimum, but only increases the curvature of the potential at the minimum. Thus, the field $\phi$ remains anchored at its vacuum and, again, the universe ends up empty when $\Lambda \approx 0$. 
 
 The only case in which inflation can kick in efficiently is when the last transition is between a configuration with unbroken EW 
  and a configuration in which EW is broken for the first time with $\Lambda \approx 0$. After this transition, $\phi$ finds itself far from its minimum and, for an appropriate functional form of $F(\phi)$, the field evolution can activate a phase of slow-roll inflation followed by reheating.
  
 In this example, the selection criterion of having a non-empty universe identifies {\it uniquely} a single universe: the one where EW is broken for the first time in an environment with nearly vanishing cosmological constant.  
 
\section{Conclusions}

The Selfish Higgs is a mechanism that selects a small cosmological constant and small Higgs mass as the only possibility to have a non-empty universe. In other words, the Selfish Higgs is an evolutionary process leading to a fairly unique universe that contains something rather than nothing.  

The theory is described by the SM with the addition of a four-form and an inflaton. The four-form is naturally coupled to the SM Higgs and leads to a landscape where the cosmological constant and the Higgs mass vary in a correlated way. 
Universes with large positive or negative cosmological constant are ruled out because they cannot support structures, as argued in ref.~\cite{Weinberg:1987dv}. Out of the universes with $\Lambda \approx 0$, only those with small $\mu_H^2$ can be populated by matter and radiation, while all others remain empty. In this paper, we have proposed two different mechanisms to achieve this situation.

In the first example, we have used particle production during the non-adiabatic process of brane nucleation. Almost all universes are empty because particle production is exponentially suppressed whenever $|\mu_H^2|$ is larger than the brane charge, which is the quantity that defines the weak scale. In these empty universes, the inflaton lies inert at its minimum. Only under the special circumstances of simultaneously small $\Lambda$ and $\mu_H^2$, a thermal bath of SM particles is created. This thermal bath acts as a match starting the fire of inflation and creating the conditions for a non-trivial universe.

The second example is based on a special form of inflationary models, in which the inflaton wanders randomly in a flat potential during the EW unbroken phase, while it is chained at its vacuum during the EW broken phase. In either case, inflation cannot take place. A special situation occurs at the interface between these two cases when, for $\Lambda \approx 0$, EW symmetry is broken for the first time in the brane nucleation process. Only in this critical situation can inflation be triggered, generating a non-empty universe. 

While it is difficult to identify model-independent experimental signatures of the Selfish Higgs, a robust conclusion is that our framework can be ruled out entirely by the observation of primordial gravitational waves through their imprint on the CMB. Such a detection would imply a very large Hubble rate during inflation, while the Selfish Higgs mechanism relies on values of $H$ of the order of the weak scale or smaller. Our mechanism also implies that the reheating temperature after inflation is at most of weak-scale size and this has implications for baryogenesis or for the relic density of new particle species. Heavy particles beyond the SM (which could be relevant for dark matter or baryogenesis) can be produced, but only through non-thermal processes. Other experimental consequences could be found, but are specific to mechanisms of low-scale inflation or special model-dependent features.

The Selfish Higgs is a framework that singles out small $\Lambda$ and $\mu_H^2$ as the only possibility to create a non-empty universe. Within this general paradigm, one can imagine many variations of the selection mechanisms, even beyond those discussed here. The Selfish Higgs is a framework that lends itself to different realisations and may offer a new perspective on the Higgs naturalness problem.

\subsection*{Acknowledgments}
We thank C.~ Bachas,  R.~Bousso, G.~Dvali, J.R.~Espinosa, M.~Geller, R.~Gregory, A.~Hook, E.~Kiritsis, A.~Linde, J.~March-Russell, M.~McCullough, A.~Strumia and N.~Tetradis  for discussions. 
A.R.~is supported by the Swiss National Science Foundation (SNSF),  project {\sl The Non-Gaussian Universe and Cosmological Symmetries}, project number: 200020-178787.  G.F.G.~thanks the Munich Institute for Astro and Particle Physics (MIAPP) of the DFG Cluster of Excellence {\sl Origin and Structure of the Universe} for hospitality during which part of this work was done.


\begin{thebibliography}{99}

\bibitem{gia1} G.~Dvali and A.~Vilenkin,
  %``Cosmic attractors and gauge hierarchy,''
  Phys.\ Rev.\ D {\bf 70}, 063501 (2004)
  \arXivold{hep-th/0304043}{}.

\bibitem{gia2} G.~Dvali,
  %``Large hierarchies from attractor vacua,''
  Phys.\ Rev.\ D {\bf 74}, 025018 (2006)
  \arXivold{hep-th/0410286}{}.


\bibitem{Graham:2015cka}
  P.~W.~Graham, D.~E.~Kaplan and S.~Rajendran,
  %``Cosmological Relaxation of the Electroweak Scale,''
  Phys.\ Rev.\ Lett.\  {\bf 115} (2015)  221801
  \arXiv{1504.07551}{hep-ph}.

\bibitem{Arkani-Hamed:2016rle}
  N.~Arkani-Hamed, T.~Cohen, R.~T.~D'Agnolo, A.~Hook, H.~D.~Kim and D.~Pinner,
  %``Solving the Hierarchy Problem at Reheating with a Large Number of Degrees of Freedom,''
  Phys.\ Rev.\ Lett.\  {\bf 117} (2016)  251801
  \arXiv{1607.06821}{hep-ph}.

\bibitem{Arvanitaki:2016xds}
  A.~Arvanitaki, S.~Dimopoulos, V.~Gorbenko, J.~Huang and K.~Van Tilburg,
  %``A small weak scale from a small cosmological constant,''
  JHEP {\bf 1705} (2017) 071
  \arXiv{1609.06320}{hep-ph}.

\bibitem{ib} A.~Herraez and L.~E.~Ibanez,
  %``An Axion-induced SM/MSSM Higgs Landscape and the Weak Gravity Conjecture,''
  JHEP {\bf 1702}, 109 (2017)
  \arXiv{1610.08836}{hep-th}.
  
\bibitem{Geller:2018xvz}
  M.~Geller, Y.~Hochberg and E.~Kuflik,
  %``Inflating to the Weak Scale,''
  Phys.\ Rev.\ Lett.\  {\bf 122} (2019)  191802
  \arXiv{1809.07338}{hep-ph}.

\bibitem{Cheung:2018xnu}
  C.~Cheung and P.~Saraswat,
  %``Mass Hierarchy and Vacuum Energy,''
  \arXiv{1811.12390}{hep-ph}.

\bibitem{dawkins}
R.~Dawkins, {\it The Selfish Gene}, Oxford University Press, 1976.

\bibitem{ANT} 
  A.~Aurilia, H.~Nicolai and P.~K.~Townsend,
  Nucl.\ Phys.\ B {\bf 176}, 509 (1980).
 % doi:10.1016/0550-3213(80)90466-6

\bibitem{Witten-4} 
E.~Witten, {\it Proceedings of the 1983 Shelter Island Conference on Quantum Field Theory and the Fundamental Problems of Physics}, 
 % E.~Witten,
  %``Fermion Quantum Numbers In Kaluza-klein Theory,''
  Conf.\ Proc.\ C {\bf 8306011}, 227 (1983).

\bibitem{HT}
  M.~Henneaux and C.~Teitelboim,
  %``The Cosmological Constant As A Canonical Variable,''
  Phys.\ Lett.\  {\bf 143B} (1984) 415.

\bibitem{BT1}
  J.~D.~Brown and C.~Teitelboim,
  %``Dynamical Neutralization of the Cosmological Constant,''
  Phys.\ Lett.\ B {\bf 195} (1987) 177.
  %%CITATION = PHLTA,B195,177;%%
  %162 citations counted in INSPIRE as of 10 juil. 2015

\bibitem{BT2}
  J.~D.~Brown and C.~Teitelboim,
 % ``Neutralization of the Cosmological Constant by Membrane Creation,''
  Nucl.\ Phys.\ B {\bf 297} (1988) 787.
  %%CITATION = NUPHA,B297,787;%%
  %196 citations counted in INSPIRE as of 10 Jul 2015

\bibitem{duff} 
  M.~J.~Duff,
 % ``The Cosmological Constant Is Possibly Zero, but the Proof Is Probably Wrong,''
  Phys.\ Lett.\ B {\bf 226}, 36 (1989)
  [Conf.\ Proc.\ C {\bf 8903131}, 403 (1989)].
  %%CITATION = PHLTA,B226,36;%%

\bibitem{BP}
  R.~Bousso and J.~Polchinski,
 % ``Quantization of four form fluxes and dynamical neutralization of the cosmological constant,''
  JHEP {\bf 0006} (2000) 006
 \arXivold{hep-th/0004134}.
 
 
 \bibitem{saltatory}
  J.~L.~Feng, J.~March-Russell, S.~Sethi and F.~Wilczek,
  %``Saltatory relaxation of the cosmological constant,''
  Nucl.\ Phys.\ B {\bf 602} (2001) 307
  \arXivold{hep-th/0005276}.
  %%CITATION = HEP-TH/0005276;%%
  %188 citations counted in INSPIRE as of 10 juil. 2015  
  
    \bibitem{Gia0} 
  G.~R.~Dvali and A.~Vilenkin,
 % ``Field theory models for variable cosmological constant,''
  Phys.\ Rev.\ D {\bf 64}, 063509 (2001)
  \arXivold{hep-th/0102142}.
  %%CITATION = HEP-TH/0102142;%%
  %27 citations counted in INSPIRE as of 15 juil. 2015  

\bibitem{FKRR} 
  F.~Farakos, A.~Kehagias, D.~Racco and A.~Riotto,
  JHEP {\bf 1606}, 120 (2016)
 % doi:10.1007/JHEP06(2016)120
  \arXiv{1605.07631}{hep-th}.  

\bibitem{Freese:2006fk}
    K.~Freese, J.~T.~Liu and D.~Spolyar,
    % ``Chain inflation via rapid tunneling in the landscape,"
   \arXivold{hep-th/0612056}.

  \bibitem{K1} 
  N.~Kaloper and L.~Sorbo,
 % ``A Natural Framework for Chaotic Inflation,''
  Phys.\ Rev.\ Lett.\  {\bf 102}, 121301 (2009)
  \arXiv{0811.1989}{hep-th}.
  %%CITATION = ARXIV:0811.1989;%%
  %123 citations counted in INSPIRE as of 15 juil. 2015  
  
  \bibitem{K2} 
  N.~Kaloper, A.~Lawrence and L.~Sorbo,
 % ``An Ignoble Approach to Large Field Inflation,''
  JCAP {\bf 1103}, 023 (2011)
  \arXiv{1101.0026}{hep-th}.
  %%CITATION = ARXIV:1101.0026;%%
  %119 citations counted in INSPIRE as of 15 Jul 2015
  
  \bibitem{K3} 
  N.~Kaloper and A.~Lawrence,
 % ``Natural chaotic inflation and ultraviolet sensitivity,''
  Phys.\ Rev.\ D {\bf 90} 023506 (2014)
  \arXiv{1404.2912}{hep-th}.
  %%CITATION = ARXIV:1404.2912;%%
  
\bibitem{Dudas}
  E.~Dudas,
 % ``Three-form multiplet and Inflation,''
  JHEP {\bf 1412} (2014) 014
  \arXiv{1407.5688}{hep-th}.
  %%CITATION = ARXIV:1407.5688;%%
  %8 citations counted in INSPIRE as of 15 Jul 2015  
  
  \bibitem{Gia1}
  G.~Dvali,
 % ``Three-form gauging of axion symmetries and gravity,''
  \arXivold{hep-th/0507215}.
  %%CITATION = HEP-TH/0507215;%%
  %23 citations counted in INSPIRE as of 15 juil. 2015
  
\bibitem{Gia2}
  G.~Dvali,
 % ``A Vacuum accumulation solution to the strong CP problem,''
  Phys.\ Rev.\ D {\bf 74} (2006) 025019
  \arXivold{hep-th/0510053}.
  
   \bibitem{Weinberg:1987dv}
  S.~Weinberg,
  %``Anthropic Bound on the Cosmological Constant,''
  Phys.\ Rev.\ Lett.\  {\bf 59} (1987) 2607.
 
  
\bibitem{Duncan:1989ug}
  M.~J.~Duncan and L.~G.~Jensen,
  %``Four Forms and the Vanishing of the Cosmological Constant,''
  Nucl.\ Phys.\ B {\bf 336} (1990) 100.


\bibitem{Gibbons:1976ue}
  G.~W.~Gibbons and S.~W.~Hawking,
  %``Action Integrals and Partition Functions in Quantum Gravity,''
  Phys.\ Rev.\ D {\bf 15} (1977) 2752.
   
  \bibitem{quevedo} S.~de Alwis, R.~Gupta, E.~Hatefi and F.~Quevedo,
 % ``Stability, Tunneling and Flux Changing de Sitter Transitions in the Large Volume String Scenario,''
  JHEP {\bf 1311}, 179 (2013)
  \arXiv{1308.1222}{hep-th}.
  
  \bibitem{instab1}
  G.~Degrassi, S.~Di Vita, J.~Elias-Mir\'o, J.R.~Espinosa, G.F.~Giudice, G.~Isidori and A.~Strumia,
  %``Higgs mass and vacuum stability in the Standard Model at NNLO,''
  JHEP {\bf 1208} (2012) 098
  \arXiv{1205.6497}{hep-ph}.
  %%CITATION = doi:10.1007/JHEP08(2012)098;%%
  
  \bibitem{instab2}  
 D.~Buttazzo, G.~Degrassi, P.P.~Giardino, G.F.~Giudice, F.~Sala, A.~Salvio and A.~Strumia,
  %``Investigating the near-criticality of the Higgs boson,''
  JHEP {\bf 1312} (2013) 089
  \arXiv{1307.3536}{hep-ph}.
  %%CITATION = doi:10.1007/JHEP12(2013)089;%%
  
    \bibitem{chen2} X.~Chen, Y.~Wang and Z.~Z.~Xianyu,
  %``Standard Model Mass Spectrum in Inflationary Universe,''
  JHEP {\bf 1704}, 058 (2017)
 \arXiv{1612.08122}{hep-th}.

  \bibitem{pt} L.~E.~Parker and D.~Toms,
  {\it Quantum Field Theory in Curved Spacetime: Quantized Field and Gravity}, Cambridge University Press, 2009.
  
  \bibitem{jump} M.~Herranen, T.~Markkanen, S.~Nurmi and A.~Rajantie,
  %``Spacetime curvature and Higgs stability after inflation,''
  Phys.\ Rev.\ Lett.\  {\bf 115}, 241301 (2015)
  \arXiv{1506.04065}{hep-ph}.

  
  \bibitem{gw} A.~H.~Guth and E.~J.~Weinberg,
%``Could the Universe Have Recovered from a Slow First Order Phase Transition?,''
  Nucl.\ Phys.\ B {\bf 212}, 321 (1983).

 
   \bibitem{Linde:1996cx}
  A.~D.~Linde,
  %``Relaxing the cosmological moduli problem,''
  Phys.\ Rev.\ D {\bf 53} (1996) R4129
   \arXivold{hep-th/9601083}{}.
   
   \bibitem{Dine:1995uk}
  M.~Dine, L.~Randall and S.~D.~Thomas,
  %``Supersymmetry breaking in the early universe,''
  Phys.\ Rev.\ Lett.\  {\bf 75} (1995) 398
  \arXivold{hep-ph/9503303}{}.
 
 \bibitem{kt} L. ~Knox and M.~S.~Turner, Phys. Rev. Lett. {\bf 70} (1993) 371
  \arXivold{astro-ph/9209006}{}.

 \bibitem{Espinosa:2015eda}
  J.~R.~Espinosa, C.~Grojean, G.~Panico, A.~Pomarol, O.~Pujol\`as and G.~Servant,
  %``Cosmological Higgs-Axion Interplay for a Naturally Small Electroweak Scale,''
  Phys.\ Rev.\ Lett.\  {\bf 115} (2015) 251803
 \arXiv{1506.09217}{hep-ph}.
 
 

\end{thebibliography}
\end{document}